\documentclass[pra,12pt,tightenlines,showpacs]{revtex4}

\usepackage{bm}
\usepackage{graphicx}

\begin{document}

\newcommand{\sgn}{\,\mbox{sgn}\,}

\title{Hamiltonian mappings and circle packing phase spaces: \\ numerical investigations}

\author{A. J. Scott}
\email{ascott@phys.unm.edu}
\affiliation{Department of Physics and Astronomy, The University of New Mexico, Albuquerque, NM 87131-1156, USA.}

\begin{abstract}
In a previous paper we introduced examples of Hamiltonian mappings with phase space structures 
resembling circle packings. We now concentrate on one particular mapping and present numerical 
evidence which supports the conjecture that the set of circular resonance islands is dense in phase 
space. 
\end{abstract}

\maketitle

\section{Introduction}\label{sectionI}

Piecewise isometries are iterated mappings which preserve distances on each of 
a number of disjoint regions of space. Such maps and their variants are currently 
enjoying considerable attention in the literature
\cite{ashwin1,ashwin2,goetz1,goetz2,goetz3,goetz4,goetz5,ashwin3,adler,kahng,ashwin4,ashwin5}, 
and arise naturally in the context of digital filters 
\cite{ebert,chua1,chua2,ogorzalek,galias,wu,kocarev1,davis,kocarev2}
and dual billiards \cite{vivaldi,tabachnikov1,tabachnikov2}. In a recent article \cite{scott1}
three new piecewise isometries were introduced, each corresponding 
to a particular phase space geometry: planar, hyperbolic or spherical. As examples 
of kicked Hamiltonian systems, all of these maps invite physical interpretations. For instance,
in the presence of damping, the planar map describes oscillations in a water-filled U-tube 
under the periodic kicking of bubbles \cite{mudde}. 

In the present paper we restrict our attention to the mapping defined on a sphere. 
It has a dynamical behaviour which is dominated by an infinite number of stable periodic 
orbits of arbitrarily long period. The resonance islands associated with these orbits 
are circular discs which fill in the phase space in a manner resembling circle packings 
\cite{herrmann,bullet,keen,parker}. 

A {\it circle packing} is a set with empty interior 
and whose complement is the union of disjoint open circular discs. Circle packings 
with zero Lebesgue measure are called {\it complete}. We have previously conjectured 
\cite{scott1} that the phase space structure produced by this map is indeed a circle packing, 
but unlikely to be complete. We now present numerical evidence which supports this belief. 
Our choice to concentrate on the spherical map was due to the compactness of its phase space. 
This proves to be of a great advantage in the numerical computations. However all three of the 
maps described in \cite{scott1} are conjectured to produce circle packings.

Numerical studies have led Ashwin {\it et al} \cite{ashwin1,ashwin2} to put forward similar 
conjectures for the sawtooth standard map (see also Goetz \cite{goetz5} for proofs). This mapping 
might be considered an analogue of those described above for the torus. However the fundamental 
question about the `packed' nature of the phase space remained untested. To this endeavour the notion of 
{\it riddling} was developed \cite{alexander,ashwin6,ashwin4}. Also of interest in the literature 
is the concept of a {\it fat fractal} \cite{farmer,umberger,grebogi,eykholt1,tricot,eykholt2,eykholt3}. Fat 
fractals are loosely defined to be sets with a fractal structure but non-zero measure.
The circle packings described above might be considered examples of such sets.

To test whether the set of circular resonances is dense in phase space, we will employ a 
symbolic labeling of the periodic orbits and use special symmetries in the mapping, 
to efficiently calculate a large class of the discs. Our numerical studies indicate that 
this class may in fact itself pack the phase space, and therefore, is the entire set. 

Our paper is organized as follows. In the next section we will give an overview of the mapping 
to be analyzed, stating three main conjectures concerning its behaviour. In section 
\ref{sectionIII} we present numerical evidence which supports these conjectures. 
Finally, in section \ref{sectionIV} we will discuss our results. 

\section{The map}\label{sectionII}

The Hamiltonian under consideration is that of a kicked linear top
\begin{equation}
H({\bf J},t)=\omega J_3+\mu|J_1|\sum_{n=-\infty}^{\infty}\delta(t-n), \label{H}
\end{equation}
where $\mu,\omega\in[0,2\pi)$ are parameters and $({\bf J})_i=J_i=\epsilon_{ijk}x_jp_k$ $(i=1,2,3)$ are 
the three components of angular momentum for a particle confined to a sphere, normalized such that 
${\bf J}\cdot{\bf J}=1$. The evolution of ${\bf J}$ is governed by the equations
\begin{equation}
\dot{J_i}=\{J_i,H\},\quad \{J_i,J_j\}=\epsilon_{ijk}J_k
\end{equation}
where $\{\cdot\,,\cdot\}$ are the Poisson brackets. Their solution can be written 
as the mapping
\begin{eqnarray}
{\bf J}^{n+1} = \left[\begin{array}{c} J_1^{n+1} \\ J_2^{n+1} \\ J_3^{n+1} \end{array}\right]  &=&  \left[\begin{array}{ccc} \cos{\omega} & -\sin{\omega} & 0 \\ \sin{\omega} & \cos{\omega} & 0 \\ 0 & 0 & 1 \end{array}\right]\left[\begin{array}{ccc} 1 & 0 & 0 \\ 0 & \cos{\mu s^n} & -\sin{\mu s^n} \\ 0 & \sin{\mu s^n} & \cos{\mu s^n} \end{array}\right]\left[\begin{array}{c} J_1^n \\ J_2^n \\ J_3^n\end{array}\right] \nonumber\\
&\equiv& \mbox{F}(s^n){\bf J}^n \equiv F{\bf J}^n \label{map}
\end{eqnarray}
where $s^n\equiv\sgn J_1^n$, which takes ${\bf J}$ from just before a kick to one 
period later. Here $\sgn x$ is the signum function with the convention $\sgn 0=0$. 

\begin{figure}[t]
\includegraphics[scale=0.65]{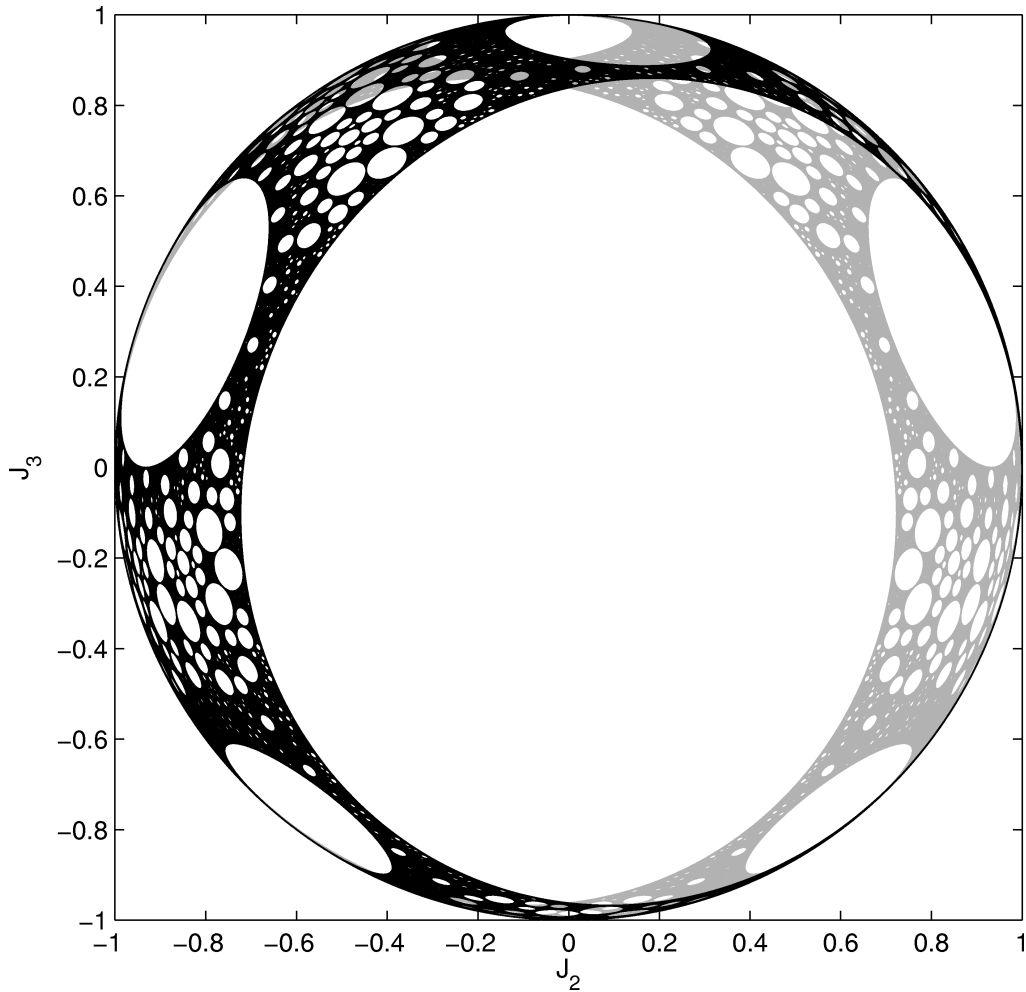}
\caption{The phase space when $\mu=\omega=\pi(\sqrt{5}-1)$.}
\label{fig1}
\end{figure}

The mapping is deceptively simple. It first rotates the eastern hemisphere ($J_1>0$) 
through an angle $\mu$ and the western hemisphere ($J_1<0$) through an angle 
$-\mu$. The great circle $J_1=0$ remains fixed. Then the entire sphere is rotated 
about the $J_3$ axis through an angle $\omega$. Thus the map rotates every point 
on the sphere in a piecewise linear fashion except on $J_1=0$ where its Jacobian is singular. 
Surprisingly, this leads to a phase space of great structural complexity. An example is 
plotted in figure \ref{fig1}. The eastern hemisphere is shown in black, the western in gray. 
To simplify our study we will restrict ourselves to the case $\mu=\omega=\pi(\sqrt{5}-1)$, 
as shown in the figure.

Note that if an orbit of our map does not have a point on $J_1=0$, then its 
Lyapunov exponents are zero, making it stable. If however, it does have a point on 
$J_1=0$, then its Lyapunov exponents are undefined. We overcome this problem by 
simply defining an unstable orbit to be one with a point on $J_1=0$, and all 
others stable. The {\it unstable set} is defined to be the closure of the set of all 
images and preimages of the great circle $J_1=0$. In figure \ref{fig1} the unstable 
set is in black and gray. The circular holes are resonances consisting entirely of 
stable orbits. At their center lies a stable periodic orbit. All other 
points inside the resonances rotate about the central periodic orbit in a 
linear fashion. The period of rotation is an irrational multiple of $\pi$ 
for almost all values of the parameters $\mu$ and $\omega$, hence their circular shape. 
If the period of rotation happens to chance upon a rational 
multiple of $\pi$, then the resonance forms a polygon (see \cite{scott1,scott2}). 

We have previously shown \cite{scott1} that it is possible to label every point of a 
periodic orbit of least period $n$ with a unique sequence $\{s^k=0,\pm 1\}_{k=0..n-1}$. 
The position of the point is given by the solution of
\begin{equation}
{\bf J}=\mbox{F}(s^{n-1})\mbox{F}(s^{n-2})\dots \mbox{F}(s^0){\bf J}\equiv\mbox{R}{\bf J}
\label{per}
\end{equation}
lying on the unit sphere with $\sgn J_1=s^0$. The other $n-1$ points of the 
periodic orbit are found by substituting the $n-1$ cycles of the original sequence 
into (\ref{per}). Hence, if $\{{\bf J}^k\}_{k=0..n-1}$ is a periodic orbit 
then the sequence corresponding to its first point is $\{\!\sgn J_1^k\}_{k=0..n-1}$. 
The orbit is unstable if and only if this sequence 
contains $0$. Note that the (unnormalized) solution of (\ref{per}) is simply the 
axis of rotation of R, namely
\begin{equation}
{\bf J}=(\mbox{R}_{23}-\mbox{R}_{32}\, ,\; \mbox{R}_{31}-\mbox{R}_{13}\, ,\; \mbox{R}_{12}-\mbox{R}_{21}). 
\end{equation}

Although every periodic point is uniquely represented by a sequence, not every 
sequence represents a periodic point. Hence we still need to find which sequences 
are legitimate. This may be done by checking each of the $2^n$ different possible 
sequences for a period-$n$ orbit. However, this is computationally expensive for 
large periods. An alternative approach is to exploit symmetries of our mapping.

If we decompose our map into involutions \cite{pina}
\begin{equation}
F=I_2\circ I_1=\tilde{I}_2\circ \tilde{I}_1
\end{equation}
where
\begin{eqnarray}
I_1{\bf J} & = & (-J_1\, ,\;J_2\cos(\mu\sgn J_1)-J_3\sin(\mu\sgn J_1)\, ,\;J_2\sin(\mu\sgn J_1)+J_3\cos(\mu\sgn J_1)) \\
I_2{\bf J} & = & (-J_1\cos{\omega}-J_2\sin{\omega}\, ,\;-J_1\sin{\omega}+J_2\cos{\omega}\, ,\;J_3) \\
\tilde{I}_1{\bf J} & = & (J_1\, ,\;-J_2\cos(\mu\sgn J_1)+J_3\sin(\mu\sgn J_1)\, ,\;J_2\sin(\mu\sgn J_1)+J_3\cos(\mu\sgn J_1)) \\
\tilde{I}_2{\bf J} & = & (J_1\cos{\omega}+J_2\sin{\omega}\, ,\;J_1\sin{\omega}-J_2\cos{\omega}\, ,\;J_3)
\end{eqnarray}
and
\begin{equation}
I_1^2=I_2^2=\tilde{I}_1^2=\tilde{I}_2^2=1
\end{equation}
we obtain the four symmetry lines 
\begin{eqnarray}
S_1 \;:\; J_1 &=& 0 \\ 
S_2 \;:\; J_1 &=& -J_2\tan(\omega/2) \\
\tilde{S}_1 \;:\; J_2 &=& J_3\tan(\mu/2)\sgn J_1 \\
\tilde{S}_2 \;:\; J_2 &=& J_1\tan(\omega/2)
\end{eqnarray}
as the fixed lines ($\text{Fix}(I)\equiv\{x|Ix=x\}$) of our involutions. Given that 
$IFI=F^{-1}$ for all the above involutions, it is not difficult to prove the following 
theorem \cite{lamb}. \\
\\
\noindent{\bf Theorem.} {\it Let $\{{\bf J}^k\}_{k=0..n-1}$ be a periodic orbit of $F$ 
with least period $n$. Suppose that the point ${\bf J}^0$ lies on a symmetry line of $F$ 
($S_1$, $S_2$, $\tilde{S}_1$ or $\tilde{S}_2$). Then no other points of the orbit lie on 
the symmetry line if $n$ is odd, or exactly one other, ${\bf J}^{n/2}$, if $n$ is even.} \\
\\  
This property reduces the task of finding periodic orbits to a one-dimensional search. 
However, if we wish to find {\it all} periodic orbits of a given period, then we must first
establish that every periodic orbit {\it must} have a point on at least one of the four 
given symmetry lines. By definition, every unstable orbit has a point on $S_1$, and thus, 
we have the following fact. \\
\\
\noindent{\bf Fact. } {\it Every unstable periodic orbit has a point on $S_1$.} \\
\\
In \cite{scott1} we made the following conjecture based on preliminary numerical work. \\ 
\\
\noindent{\bf Conjecture 1.} {\it For almost all $\omega$ and $\mu$, every stable periodic 
orbit has a point on at least one of $S_2$, $\tilde{S}_1$ or $\tilde{S}_2$.} \\
\\
The conjecture is not true for all $\omega$ and $\mu$ since, as remarked above, 
the rotation of a resonance may become a rational multiple of $\pi$ when the 
parameters take on particular values. However this occurs on a set of measure zero.  

The above theorem, and the corresponding {\it symmetric} periodic orbits for which it 
refers, have become folklore in the theory of reversible dynamical systems \cite{lamb}. 
Proponents of symmetry methods include Birkoff \cite{birkoff}, De Vogelaere \cite{devogelaere}
and Greene {\it et al.} \cite{greene}. Unfortunately, one generally finds that many 
other periodic orbits will be present which do not have points on a symmetry line. Thus, 
the mapping we are considering is conjectured to be of a very special class for which 
{\it all} periodic orbits are of the symmetric type.

We now turn our attention to the unstable set. The complicated structure of circular 
resonances with decreasing radii displayed in figure \ref{fig1} suggests the following 
conjecture. \\
\\
\noindent{\bf Conjecture 2.} {\it The unstable set has empty interior.} \\
\\
That is, the unstable set is a {\it circle packing} of the unit sphere. We also put forward 
a third conjecture. \\
\\
\noindent{\bf Conjecture 3.} {\it The unstable set has positive Lebesgue measure.} \\
\\
Or equivalently, the circle packing is {\it not} complete. Numerical evidence which 
supports the above three conjectures will now be presented.

\section{Numerical evidence}\label{sectionIII}

Suppose we ignore the fact that the stable symmetric periodic orbits may only form a 
subset of the set of all stable periodic orbits. If we find that the set of circular 
resonances associated with these orbits {\it pack} the phase space (i.e. the set of 
resonance discs is dense on the surface of the unit sphere), then we are left with only 
one conclusion: both Conjecture 1 and 2 must be correct. Furthermore, we are now in a 
position to test Conjecture 3. This is precisely the approach we will take. Consider the 
following trivial property. 

Suppose ${\bf J}^0$ lies on a symmetry line corresponding to the involution $I$ 
($I{\bf J}^0={\bf J}^0$), and the iterate ${\bf J}^m$ also lies on this same 
symmetry line ($IF^m{\bf J}^0=F^m{\bf J}^0$). Then, 
\begin{equation}
F^{2m}{\bf J}^0=F^mF^m{\bf J}^0=F^mIF^m{\bf J}^0=IF^{-m}I^2F^m{\bf J}^0=I{\bf J}^0={\bf J}^0
\end{equation}
and hence, ${\bf J}^0$ must be a periodic point with period $2m$. 
To find the orbit's {\it least} period we construct the unique sequence 
$\{\!\sgn J_1^k\}_{k=0..2m-1}$ and locate the shortest {\it word} which 
generates the orbit. If ${\bf J}^m$ was the first iterate to hit the symmetry line, 
then either the period of the orbit is odd with period $m$, or even with period $2m$.  

Thus we only need to find where each symmetry line intersects with its image; an effortless 
task to solve numerically by virtue of the simplicity of our mapping. Every image of a 
symmetry line is just a collection of arc segments on the sphere.

Now suppose that we have just found a particular periodic orbit. If we are to gain some 
insight into the above three conjectures we will need to find the size of its 
circular resonances. Note that the resonance of at least one point of the orbit must 
touch the great circle $J_1=0$ ($S_1$). This is because the boundary of each resonance 
forms part of the unstable set, and hence, iterates arbitrarily close to $S_1$. 
Consequently 
\begin{equation}
\sigma=\min_{k=0..n-1} |J_1^k|
\end{equation}
where $n$ is the period of the orbit, and $\sigma$ is perpendicular distance between the 
boundary of the circular resonance, and the line connecting its center (the periodic point)
with the origin. All resonances of the same periodic orbit are of equal size. The sections 
of the symmetry line which intersect the orbit's resonances contain no other 
periodic orbits and may be ignored for the remainder of our search.

For the symmetry lines $S_2$, $\tilde{S}_1$ and $\tilde{S}_2$, the above procedure was 
run for $m=5\times 10^5$ iterations of the mapping. Hence all symmetric periodic 
orbits with odd periods $\leq 5\times 10^5$ and even periods $\leq 1\times 10^6$ were 
found, producing over 22 billion circles. 

To investigate properties of the unstable set we will define the radius of a circle 
on the unit sphere to be 
\begin{equation}
r \equiv 2\sin(\arcsin(\sigma)/2)
\end{equation} 
which allows us to write the spherical area covered by the resonance as $\pi r^2$.  
Other definitions for the radius such as $\sigma$, or the angular radius $\arcsin(\sigma)$, 
are also suitable since all are equivalent in the limit $\sigma\rightarrow 0$. 
\begin{figure}[t]
\includegraphics[scale=0.9]{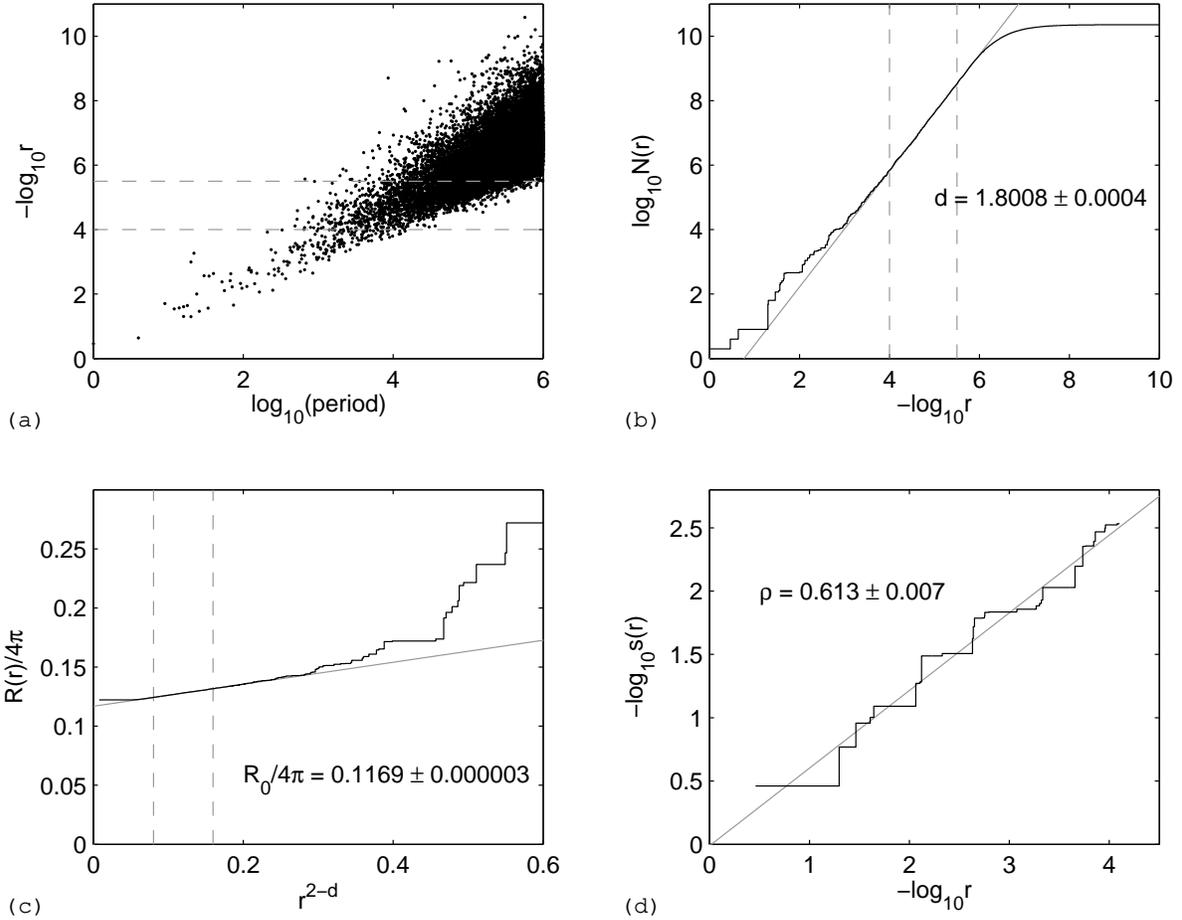}
\caption{(a) The radii of the circles for increasing periods of the orbit. (b) The number 
of circles with radius $>r$. (c) The residual area not covered by the set of circles 
with radius $>r$. (d) The radius of the largest circle which will fit into the complement 
of this set.}
\label{fig2}
\end{figure}

The proliferation of circles with smaller and smaller radii is displayed in figure 
\ref{fig2}(a). Here the radius of the resonance versus the period of the orbit is 
plotted on a logarithmic scale. Note that we can only be sure that we have all circles 
of a particular radius for radii $\lesssim 10^{-5.5}$ (the upper dashed line in figure 
\ref{fig2}(a)). 

Next we define the quantity $N(r)$ as the number of circles of radius 
larger than $r$. It is found that $N(r)$ follows the scaling law
\begin{equation}
N(r)=Ar^{-d}
\end{equation}
(see figure \ref{fig2}(b)). Linear regression on a data set in the range 
$10^{-4}\leq r\leq 10^{-5.5}$ reveals the exponent to be $d\approx 1.8$. If the 
unstable set is indeed a circle packing, then the above scaling law implies that $d$ 
will be its {\it circle packing exponent}
\begin{equation}
d=\sup\bigg\{x:\sum_{j=1}^{\infty}r_j^x=\infty\bigg\}=\inf\bigg\{y:\sum_{j=1}^{\infty}r_j^y<\infty\bigg\}
\end{equation}
where the radii $r_j$ are labeled to form a non-increasing sequence: 
$r_1\geq r_2\geq\dots.$ The circle packing exponent is known as the 
Besicovitch-Taylor index in one dimension \cite{besicovitch}, and is equivalent to the 
exterior dimension $d_x$ and uncertainty exponent $\alpha$ 
of Grebogi {\it et al} \cite{grebogi}: $d=d_x=2-\alpha$ \cite{tricot}. 

To investigate the Lebesgue measure of the unstable set we define the residual area
\begin{equation}
R(r)\equiv 4\pi-\pi\sum_{j=1}^{j(r)}r_j^2
\end{equation}
where $j(r)$ is the largest $j$ such that $r_j>r$. This quantity follows the scaling 
law 
\begin{equation}
R(r)=R_0+Br^{2-d}.
\end{equation}
Using linear regression on our data set (see figure \ref{fig2}(c)) we find that 
$R_0/4\pi\approx0.1169$, and hence, the set of resonances (for the symmetric periodic 
orbits) measure less than $89\%$ of the sphere. Note that the above scaling law also implies
that $2-d=\beta$, where $\beta$ is the fatness exponent of Farmer \cite{farmer}.

Finally we ask the most critical question: Does the set of resonances for the 
symmetric periodic orbits {\it pack} the surface of the sphere? A numerical scheme 
for answering this question might be to consider the set $C_r$ of all circular resonances with 
radius larger than $r$. If we define the quantity $s(r)$ to be the radius of the largest 
circle which will `fit' into the complement of this set, then the 
unstable set has empty interior if and only if $s(r)\rightarrow 0$ as $r\rightarrow 0$. 
In figure \ref{fig2}(d) we plot $s(r)$ on a logarithmic scale and find that the scaling law 
\begin{equation}
s(r)=Cr^{\rho}
\end{equation}
is approximately obeyed. The exponent might be thought of as describing the `completeness' 
of the circle packing, and invites the general definition 
\begin{equation}
\rho\equiv \lim_{r\rightarrow 0}\frac{\log s(r)}{\log r}. \label{rho}
\end{equation} 
When constructing Apollonian packings, $s(r)$ is the radius of the next circle to be 
packed, and hence $\rho=1$. In our case $\rho\approx 0.6$. This estimate, however, might 
prove to be a poor one. Note that $s(r)$ has been calculated only for $r\lesssim 10^{-4}$. 
Ideally we would like to use our entire set of circles and increase the graph to 
$r=10^{-5.5}$. However the current plot took over a year to produce. Our procedure 
for finding all symmetric circles with radius larger than $r$ is extremely efficient. 
However we are unaware of an efficient algorithm for the calculation of $s(r)$. 
 
\section{Conclusion}\label{sectionIV}

In this paper we have outlined a procedure to efficiently calculate all symmetric periodic orbits 
up to a given period for the spherical map of \cite{scott1}. When $\mu=\omega=\pi(\sqrt{5}-1)$, 
the circular resonance islands corresponding to the stable orbits are found to initially follow a 
scaling law which indicates that they `pack' the surface of the unit sphere. This would imply that 
{\it all} periodic orbits are of the symmetric type. Under this assumption, the unstable set 
has empty interior, but appears not to be of zero Lebesgue measure. The latter result agrees with 
Ashwin {\it et al} \cite{ashwin1,ashwin2} for the sawtooth standard map. As a final remark, 
it might prove fruitful to study the index $\rho$ (\ref{rho}) as an indicator for nowhere dense sets.  
Measures of connectedness have already been defined \cite{robins}.

\begin{acknowledgments}
The author would like to thank Vanessa Robins for helpful discussions. This work was supported by 
the Office of Naval Research Grant No. N00014-00-1-0578.
\end{acknowledgments}

\end{document}